\documentclass[prb,twocolumn,superscriptaddress,showpacs,floatfix]{revtex4-1}  
\usepackage{amsmath}
\usepackage{color}
\usepackage{txfonts}
\usepackage{mathrsfs}
\usepackage{dcolumn}
\usepackage{bm}
\usepackage{graphicx}
\usepackage{multirow}

\usepackage[pdfborder={0 0 0}]{hyperref}
\allowdisplaybreaks
\hypersetup{
    colorlinks,
    linkcolor=red,
    urlcolor=black,
    citecolor=red,
}

\begin{document}

\title{Towards ab initio self-energy embedding theory in quantum chemistry}
\date{\today}
\author{Tran Nguyen Lan}
\email{latran@umich.edu}
\altaffiliation{On leave from: Ho Chi Minh City Institute of Physics, VAST, Vietnam}
\affiliation{Department of Chemistry, University of Michigan, Ann Arbor, Michigan, 48109, USA}
\affiliation{Department of Physics, University of Michigan, Ann Arbor, Michigan, 48109, USA}
\author{Alexei A. Kananenka}
\affiliation{Department of Chemistry, University of Michigan, Ann Arbor, Michigan, 48109, USA}
\author{Dominika Zgid}
\affiliation{Department of Chemistry, University of Michigan, Ann Arbor, Michigan, 48109, USA}

\begin{abstract}
  The self-energy embedding theory (SEET), in which the active space self-energy  is embedded in the self-energy obtained from a perturbative method treating the non-local correlation effects, was recently developed in our group.
  In SEET the double counting problem does not appear and the accuracy can be improved either by increasing the perturbation order or by enlarging the active space.
  This method was first calibrated for the 2D Hubbard lattice showing promising results.
  In this paper, we report an extension of SEET to quantum chemical {\it ab initio} Hamiltonians for applications to molecular systems.
  The self-consistent second-order Green's function (GF2) method is used to describe the non-local correlations, while the full configuration interaction (FCI) method is carried out to capture strong correlation within the active space.
  Using few proof-of-concept examples, we show that SEET yields results of comparable quality to $n-$electron valence state second-order perturbation theory (NEVPT2) with the same active space, and furthermore, the full active space can be split into smaller active spaces without further implementation.
  Moreover, SEET avoids intruder states and  does not require any high-order reduced density matrices.
  These advantages show that SEET is a promising method to describe physical and chemical properties of challenging molecules requiring large active spaces.
\end{abstract}

\maketitle

{\it Introduction.}
Strongly correlated systems such as materials and molecules containing transition metal atoms present a significant challenge for both quantum chemistry and condensed matter physics.
The quantitative description of these systems is difficult since both the strong (static) 
and weak (dynamical) correlations present between electrons have to be included.

In quantum chemistry, multi-reference perturbation theories, namely complete active space second-order perturbation theory \cite{andersson1990second, ghigo2004modified} (CASPT2)
and $n-$electron valence state second-order perturbation theory \cite{angeli2001introduction, angeli2002n} (NEVPT2),
are commonly employed to simultaneously handle both types of correlation yielding quantitative accuracy without any adjustable parameters. 
Nevertheless, these methods are prohibitively expensive for any extended systems.
Conventionally, the largest computationally accessible active space  has up to 16 electrons in 16 orbitals, thus treating only very few strongly correlated orbitals in the whole system.
Recently, significant efforts have been devoted to enlarge the active space using DMRG \cite{white1999ab, chan2009density, zgid2008density, kurashige2009high, marti2011new}, RAS \cite{malmqvist1990restricted, malmqvist2008restricted}, SplitCAS \cite{manni2011strong, li2013splitgas}, and GAS \cite{ma2011generalized, vogiatzis2015systematic} techniques.
However, in these methods evaluating and storing intermediates involving 3- and 4-body reduced density matrices (RDMs) is extremely demanding as the number of active orbitals increases.
A different class of quantum chemistry approaches that simultaneously treats static and dynamical correlations combines multi-configurational methods and density functional theory 
(MC-DFT)\cite{grafenstein2000combination, grafenstein2005development, savin_mc_dft_2002, li2014multiconfiguration, carlson2015multiconfiguration}.
However, some of these methods suffer  from the double counting \cite{grafenstein2000combination}  of correlation and lack of systematic approaches to improve the DFT part.

In condensed matter, the dynamical mean-field theory (DMFT), formulated in the language of many-body Green's function,
has been extensively used to deal with the strongly correlated systems \cite{georges1996dynamical, kotliar2006electronic, kotliar2004strongly, held2007electronic}.
In DMFT, only local correlations are treated, whereas non-local correlations are neglected.
The combination of the DFT and DMFT methods called LDA+DMFT was developed to include non-local correlation between unit cells \cite{kotliar2006electronic, lechermann2006dynamical},
and it has been widely used to describe the strongly correlated materials, for instance, see Refs. \onlinecite{lichtenstein2001finite, lechermann2006dynamical, ren2006lda+}.
LDA+DMFT has been also applied to molecular systems, such as H$_2$ molecule \cite{lee2015dynamical} and transition metal complexes \cite{boukhvalov2008correlation, weber2014renormalization, weber2013importance, PhysRevB.91.241111, PhysRevB.92.035146}.
Similar to MC-DFT methods, LDA+DMFT suffers from the double counting problem and the DFT part cannot be systematically improved.
Moreover, in LDA+DMFT the impurity Hamiltonian is frequently parameterized using empirical on-site Coulomb interactions that are freely adjusted to fit experiments~\cite{vaugier2012hubbard}. 
Other diagrammatic methods, such as GW+DMFT \cite{biermann2003first, tomczak2012combined} and FLEX+DMFT (here, FLEX stands for fluctuation exchange) \cite{PhysRevB.92.085104}, 
were recently developed but their application to molecular systems has not yet been established.
Moreover, it is fair to say that all these methods rely on mapping of the system of interest to the effective low energy Hamiltonian that is then subsequently solved by the DMFT method.
If a high quantitative accuracy is desired such procedures may become problematic since the mapping onto an effective model can introduce uncontrolled errors.

Recently, we have developed a general multi-scale framework, called the self-energy embedding theory (SEET), in which the self-energy describing strongly correlated orbitals (active orbitals) is self-consistently embedded into the self-energy obtained from a method able to treat the non-local correlation effects \cite{kananenka2015systematically}.
SEET employing the full configuration interaction (FCI) method to describe few strongly correlated orbitals embedded in the self-energy obtained from the self-consistent second-order Green's function method (GF2) was first illustrated and calibrated for the 2D Hubbard lattice showing promising results.

In this paper, we generalize the SEET method to molecular quantum chemical  {\it ab initio} Hamiltonians. 
The GF2 method is used to describe the non-local correlations, while strong correlations within the active space are captured by the FCI method.
To maintain consistency with our previous work, we denote the method as SEET(FCI/GF2), where FCI/GF2 stands for the methods used to describe the strongly/weakly correlated orbitals.
In this paper, we demonstrate that the main advantages of SEET are 
{\bf (i)} Green's function language giving access to the energy as well as spectroscopic quantities such as photoelectron spectrum, 
{\bf (ii)} diagrammatic formulation allowing for the exact double counting removal,
 {\bf (iii)} systematic improvability, 
 {\bf (iv)} simultaneous treatment of multiple active spaces (similarly to multiple impurities in condensed matter problems),
 {\bf (v)} no need for any high-order density matrix in the active space.  

{\it Theory.}
In this section, we will present SEET generalization to molecular {\it ab initio} Hamiltonians.

In the first step of SEET, starting either from the HF or DFT Green's function, we perform self-consistent GF2 calculation\cite{phillips2014communication, phillips2015fractional} in the AO basis for the whole molecule.
At convergence, the second-order self-energy in the time domain reads,
\begin{align}
  \left[\Sigma^{GF2}_{mol}(i\tau)\right]_{ij} = &- \sum_{klmnpq}\left[G^{GF2}_{mol}(i\tau)\right]_{kl} \
                                                \left[G^{GF2}_{mol}(i\tau)\right]_{mn} \left[G^{GF2}_{mol}(-i\tau)\right]_{pq} \nonumber \\
                                             &\times v_{ikmq} \left( 2v_{ljpn} - v_{pjln} \right),  \label{eq:siggf2}
\end{align}
where $\textbf{G}^{GF2}_{mol}(i\tau)$ is the Green's function in the time domain and $v_{ijkl} = \int d\textbf{r}_1 d\textbf{r}_2 \phi^*_i(\textbf{r}_1)\phi_j(\textbf{r}_1) v(\textbf{r}_1-\textbf{r}_2) \phi^*_k(\textbf{r}_2)\phi_l(\textbf{r}_2)$ are 2-electron integrals in an AO basis.
$\boldsymbol{\Sigma}^{GF2}_{mol}(i\tau)$ is then transformed to the $i\omega$ imaginary frequency domain using the Fourier transformation.  
The 1-body density matrix $\textbf{P}$ is directly evaluated using the converged GF2 Green's function, $\textbf{P} = - 2 \textbf{G}^{GF2}_{mol}(i\tau = \beta)$ with $\beta=1/(k_BT)$ as inverse temperature.
This density matrix is then diagonalized to obtain natural orbitals (NOs) and occupation numbers.
Active orbitals are then chosen from this set of NOs, as is done in traditional CAS type methods.
Thus, at the GF2 convergence, we obtain a set of active orbitals and $\boldsymbol{\Sigma}^{GF2}_{mol}(i\omega)$ with a corresponding Green's function $\textbf{G}^{GF2}_{mol}(i\omega)$ both transformed to the NO basis.

After choosing active orbitals, we set up an impurity problem for these orbitals as
\begin{align} \label{imp}
[G_{mol}(i\omega)]_{act}= \left[ \left(i\omega+\mu\right)\textbf{1} - \textbf{f}^{\ nodc}_{act} -\boldsymbol{\Delta}(i\omega)-\boldsymbol{\Sigma}_{mol}(i\omega) \right]^{-1},
\end{align}
where all the matrices are in the NO basis and subscript $act$ stands for a subset of active orbitals. 
In the first iteration of SEET, $[\textbf{G}_{mol}(i\omega)]_{act}=[\textbf{G}^{GF2}_{mol}(i\omega)]_{act}$. 
$\mu$ is the chemical potential.
The hybridization $\boldsymbol{\Delta}(i\omega)$ describes coupling of the active orbitals to the remaining weakly correlated ones and the impurity Hamiltonian made out of active and bath orbitals can be written as
\begin{align}
  H_{act+bath} = H_{act} + \sum_{ub} V_{ub}\left(a_u^+a_b + a_b^+a_u\right) + \sum_{b} \epsilon_b a_b^+a_b,
\end{align}
where $H_{act}$ is the full Hamiltonian within the active space,
\begin{align}
 H_{act} =  \sum_{uv} f_{uv}^{nodc} + \frac{1}{2}\sum_{uvtw} v_{uvtw} a_u^+ a_t^+ a_v a_w,
\end{align}
where $u, v, t, w, ...$ indices describe active orbitals and  $b$ index is used for bath orbitals.
$f_{uv}^{nodc}$ is the molecular Fock matrix, 
in which the local Fock matrix in the active space is exactly subtracted to eliminates the double counting between the mean-field and exact (FCI) treatment.
The couplings $V$ and the orbital energies $\epsilon$ are fitted to the hybridization $\boldsymbol{\Delta}(i\omega)$ between active space and non-interacting bath.
For a given hybridization $\boldsymbol{\Delta}(i\omega)$, the active space Green's function and self-energy, $\textbf{G}^{FCI}_{act}(i\omega)$ and $\boldsymbol{\Sigma}^{FCI}_{act}(i\omega)$,
are evaluated using the FCI solver.
Subsequently, the hybridization is updated and SEET iterations which are DMFT-like are performed until convergence, for details see Refs.~\onlinecite{zgid2011dynamical, kananenka2015systematically}.

Let us focus now on the molecular self-energy from Eq.~\ref{imp} that is constructed as
\begin{align}
  \boldsymbol{\Sigma}_{mol}(i\omega) = \boldsymbol{\Sigma}^{GF2}_{non-local}(i\omega) + \boldsymbol{\Sigma}^{FCI}_{act}(i\omega), 
\end{align}
where the non-local weakly correlated part of the GF2 self-energy $\boldsymbol{\Sigma}^{GF2}_{non-local}(i\omega)$ is a difference between the GF2 self-energy of the whole molecule and that of the active, strongly correlated (local part),
 $ \boldsymbol{\Sigma}^{GF2}_{non-local}(i\omega) = \boldsymbol{\Sigma}^{GF2}_{mol}(i\omega) - \boldsymbol{\Sigma}^{GF2}_{act}(i\omega)$.
 The $ \boldsymbol{\Sigma}^{GF2}_{non-local}(i\omega)$ term stands for an effective many-body field experienced by the strongly correlated electrons in the active space.
 The presence of this term eliminates the need for effective $U$ integrals in the active space since all the non-local interactions between the active and remaining orbitals are described by the non-local self-energy term. 
In general,  the embedding self-energy $ \boldsymbol{\Sigma}^{GF2}_{non-local}(i\omega)$ that comes from GF2 should be updated after all the DMFT-like iterations involving FCI Green's function solver are finished. 
In this paper, however, we have chosen not to do so and we performed a self-consistent GF2 procedure only once followed by the iterations updating FCI Green's function. 

Generally, SEET(FCI/GF2) consists of two levels of theory: perturbation and diagonalization.
This is, from the theoretical point of view, similar to multi-reference second-order perturbation theories.
It is therefore worth doing the comparison between SEET(FCI/GF2) and multi-reference second-order perturbation theories, namely CASPT2/NEVPT2, as summarized in Table \ref{SEET_CAS_comparison}.

{\tiny
  \begin{table*}
  \normalsize
  \caption{\label{SEET_CAS_comparison} \normalsize Comparison between NEVPT2/CASPT2 and SEET(FCI/GF2) methods.}
  \begin{ruledtabular}
    \begin{tabular}{llllll}
      &\,\,\,\,\,\,\,\,\, NEVPT2/CASPT2                                    &     & \,\,\,\,\,\,\,\,\,SEET(FCI/GF2)  \\
      \hline                                                                    
      &-- Perturbation on top of diagonalization                          &     & -- Diagonalization (impurity solver) on top of perturbation \\
      &-- Depends on the 0th-order $\hat{H}$ and the 1st-order CI space   &     & -- Independent of  the 0th-order $\hat{H}$ ,1st-order CI space not required\\
      &-- Perturbation depends on the diagonalization step                &     & -- Perturbation and diagonalization are implemented separately \\
      &-- Single active space                                             &     & -- Multiple active spaces \\
      &-- Requires 1-, 2-, 3-, 4-RDMs for perturbation                    &     & -- Requires only 1-body Green's function for perturbation \\
      &-- Intruder states in CASPT2                                       &     & -- No intruder states \\
      &-- PT2 only describes dynamical correlation                        &     & -- GF2 partially captures strong correlation \\
      &-- Frequency independent                                        &     & -- Frequency dependence (spectroscopic quantities) \\
      &-- No convergence in frequency grid is required                    &     & -- Requires convergence in frequency grid \\
      &-- No bath fitting procedure                                       &     & -- Requires bath fitting procedure in impurity solver\\
    \end{tabular}
  \end{ruledtabular}
  \end{table*}
}

{\it Results.}
We report few proof-of-concept examples to show that our SEET(FCI/GF2) theory is applicable to quantum chemistry.
Unless otherwise noted, the ORCA program \cite{neese2012orca} was used for all calculations using standard methods (e.g., MP2, CASSCF, NEVPT2, and FCI).
The local modified DALTON code \cite{aidas2014dalton} was used to generate RHF input necessary for GF2 and to evaluate FCI active space Green's functions~\cite{zgid_chan_solver_prb}.
We use different $\beta$ and number of frequencies for different geometries to converge (in frequency grid) the electronic energy to $10^{-4}$ a.u.

\begin{figure}[!ht]
\begin{center}
\includegraphics[width=8cm,height=5.5cm]{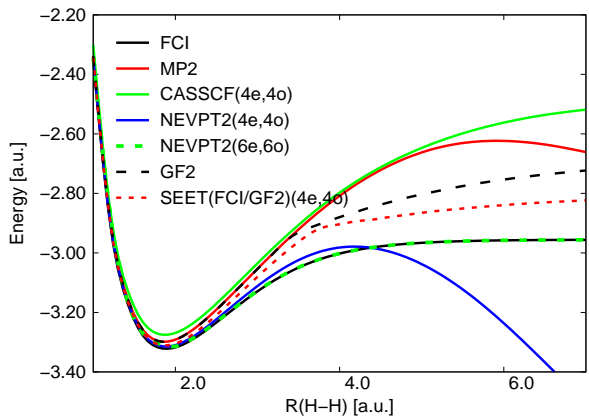}
\caption{\label{fig:h6} \normalsize
Potential energy curve of H$_6$ ring with TZ basis.} 
\end{center}
\end{figure}

In Fig.~\ref{fig:h6}, we present results for the H$_6$ ring dissociation in a TZ basis \cite{dunning1971gaussian}.
MP2 and GF2 appear identical around the equilibrium in the single-reference regime.
Upon bond stretching, the multi-reference character leads to a separation of GF2 away from MP2.
In the stretched regime, the full active space requires 6 electrons in 6 $\sigma$-type orbitals,
consequently, NEVPT2(6e,6o) agrees very well with FCI.
When smaller active space is used, CASSCF(4e,4o) incorrectly describes the dissociation and  NEVPT2(4e,4o) diverges beyond the equilibrium due to missing static correlation.
Interestingly, although SEET(FCI/GF2)(4e,4o) is very close to NEVPT2(4e,4o) at the equilibrium, it does not diverge and remains nearly parallel to FCI at long distances.
This is because GF2 itself partially recovers the static correlation~\cite{phillips2014communication} missed in NEVPT2 when the small active space is used.

\begin{figure}[!ht]
\begin{center}
\includegraphics[width=8cm,height=5.5cm]{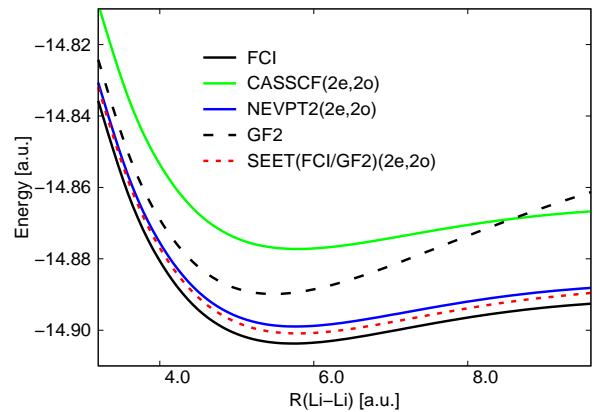}
\caption{\label{fig:li2} \normalsize
Potential energy curve of Li$_2$ with TZ basis.}
\end{center}
\end{figure}

In Fig.~\ref{fig:li2}, we consider the potential energy curve of Li$_2$ molecule in a TZ basis.
Both CASSCF(2e,2o) and NEVPT2(2e,2o) yield curves parallel to FCI.
Although GF2 yields lower energies than CASSCF(2e,2o) around the minimum, for stretched geometries GF2 is not parallel to the FCI curve.
When static correlation is properly treated by using active space on top of GF2, i.e. SEET(FCI/GF2)(2e,2o),
the dissociation is correctly described and the curve falls between NEVPT2(2e,2o) and FCI curves.

\begin{figure}[!ht]
\begin{center}
\includegraphics[width=8cm,height=5.5cm]{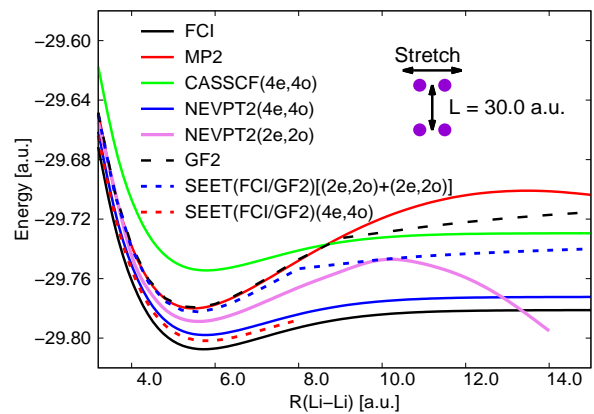}
\caption{\label{fig:li4} \normalsize
Potential energy curve of Li$_4$ with TZ basis. The inset shows the geometry of Li$_4$ cluster that includes two Li dimers separated by a fixed distance $L = 30.0$ a.u.} 
\end{center}
\end{figure}

Let us show now that SEET allows us to split the full active space into smaller active spaces composed of molecular orbitals (MOs) belonging to particular fragments.
We consider the Li$_4$ cluster made from two parallel Li dimers as shown in the inset of Fig.~\ref{fig:li4}.
The distance between these two dimers is fixed and kept long enough to avoid any coupling between them.
The cluster is stretched following the parallel direction as shown in the inset of Fig.~\ref{fig:li4}.
GF2 and MP2 yield too deep dissociation curves and as expected both curves separate when the distance is larger than 9.0 a.u.
In this system, the full active space comprises two pairs of $\sigma$-type MOs.
Consequently, both CASSCF(4e,4o) and NEVPT2(4e,4o) methods yield correct dissociation curves within a given basis set,
whereas NEVPT2(2e,2o) with only one pair of MOs in active space diverges to $-\infty$ at $R > 10.0$ a.u.
The SEET(FCI/GF2)(4e,4o) curve remains between NEVPT2(4e,4o) and FCI curves, similarly to the case of Li$_2$.
At this stage, we cannot go beyond $R = 8.0 $ a.u. for SEET(FCI/GF2)(4e,4o) calculation due to problems with convergence.

In SEET, the full active space of two pairs of MOs can be split into two smaller equivalent active spaces, where each active space includes one pair of MOs.
As shown in Fig.~\ref{fig:li4}, SEET(FCI/GF2)[(2e,2o)+(2e,2o)] dissociation curve does not diverge and remains nearly parallel to the FCI one at long distances.
While for the example of two parallel Li dimers, it is possible to localize orbitals on each fragment before splitting the full active space (similarly to the active space decomposition (ASD) developed by Parker and coworkers \cite{parker2013communication}),
we avoid doing so since we aim to demonstrate that despite missing many CI configurations when the active space consisting of MOs of the same symmetry (i.e. $\sigma-$type MO) is split, SEET can avoid divergences and recover a dissociation limit parallel to the FCI curve.
Moreover, it is evident that the SEET description can be systematically improved when enlarging the active space.

\begin{figure}[!ht]
\begin{center}
\includegraphics[width=8cm,height=5.5cm]{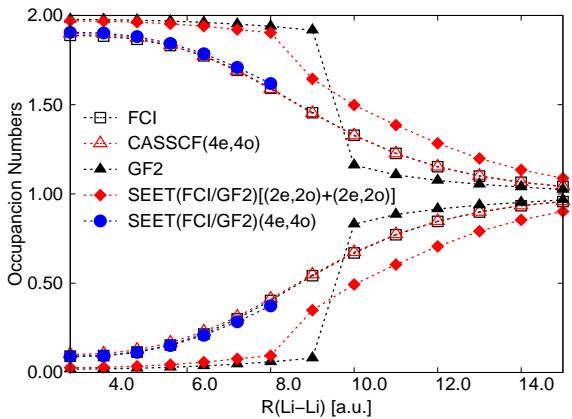}
\caption{\label{fig:li4_occ} \normalsize
Occupation numbers of Li$_4$ with respect to Li-Li distance.}
\end{center}
\end{figure}

To further explore active space splitting, in Fig.~\ref{fig:li4_occ}, as a function of bond stretching in Li$_4$ cluster, we plot orbital occupations of valence MOs obtained by FCI, CASSCF, GF2, and SEET(FCI/GF2) methods.
FCI occupations smoothly shift from single-reference to multi-reference as the bond length increases.
CASSCF(4e,4o) and SEET(FCI/GF2)(4e,4o) occupations are in a very good agreement with FCI reference.
In GF2, occupations suddenly jump from single-reference to multi-reference regime at $R = 9.0$ a.u.
This is reflected by a kink in the GF2 dissociation curve, see Fig.~\ref{fig:li4}.
Furthermore, beyond this point, GF2 occupation numbers are even closer to 1.0 than those of FCI,
indicating that GF2 overestimates static correlation in the Li$_4$ cluster.
On the other hand, when performing FCI in two active spaces on top of GF2, namely SEET(FCI/GF2)[(2e,2o)+(2e,2o)],
the single-reference to multi-reference transition becomes much smoother
indicating that SEET(FCI/GF2)[(2e,2o)+(2e,2o)] describes the correlations in a more balanced way than GF2 by itself.
Consequently, the active space splitting in SEET(FCI/GF2)  can  be used to qualitatively describe the dissociation regime.

\begin{figure}[!ht]
\begin{center}
\includegraphics[width=8cm,height=5.5cm]{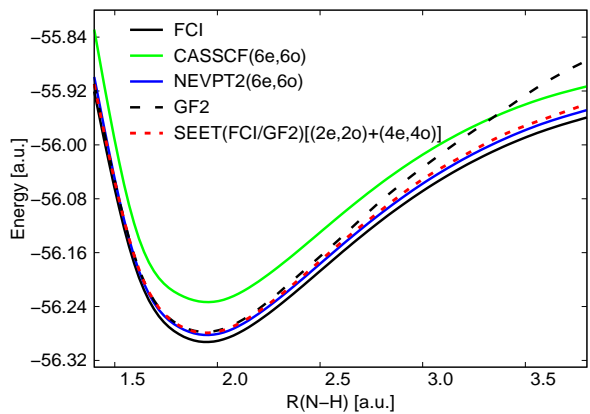}
\caption{\label{fig:nh3} \normalsize
Potential energy curve of NH$_3$ with 6-31G basis.} 
\end{center}
\end{figure}

Finally, we show that the full active space in SEET(FCI/GF2) can be split into smaller groups, where each group consists of different symmetry MOs.
To this end, we consider NH$_3$ molecule in the 6-31G basis \cite{hehre1972self}.
Fig.~\ref{fig:nh3} displays the dissociation curves from FCI, CASSCF, NEVPT2, GF2, and SEET(FCI/GF2) calculations.
For NH$_3$ molecule, the full active space is composed of 4 $\pi-$type and 2 $\sigma-$type orbitals.
Both CASSCF(6e,6o) and NEVPT2(6e,6o) correctly reproduce the FCI dissociation behavior.
Although GF2 yields the energy that is comparable to NEVPT2(6e,6o) energy at the equilibrium, it significantly differs from NEVPT2(6e,6o) at longer distances.
In the SEET(FCI/GF2) method, the full active space is split into two smaller active spaces with different orbital symmetries.
One group consists of 2 $\sigma-$type orbitals and 4 $\pi-$type orbitals are included in the other one.
It is evident that the SEET(FCI/GF2)[(2e,2o)+(4e,4o)] curve is close to that of NEVPT2(6e,6o) within the range of distances considered.
This stands in contrast to the GF2 behavior which has large error for stretched geometries.
Let us point out that in conventional CAS methods, it is also possible to split the full active space into smaller active spaces with the different orbital symmetries \cite{manni2011strong,li2013splitgas}; however, such a procedure requires a complicated implementation.
In SEET, the active space splitting does not require any additional implementation.

In conclusion, we have presented a generalization of  the SEET method to {\it ab initio} Hamiltonians for molecular systems.
GF2 and FCI were used to treat correlations in non-local (weakly correlated) and local (strongly correlated) subspaces, respectively, in a perturb and diagonalize type of scheme.
The performance of SEET(FCI/GF2) was illustrated using small molecules in small basis sets.
We demonstrated that SEET(FCI/GF2) provides results of comparable quality to NEVPT2 with the same active space.
Additionally, unlike conventional multi-reference perturbation theories, SEET avoids intruder states and  does not require high-order RDMs,
and furthermore, the full active space can be split into smaller active spaces without any additional implementation.
In contrast to LDA+DMFT type schemes, the double counting problem does not appear in SEET and the accuracy can be improved either by increasing the order of the perturbation or by enlarging the active space.
Since the non-local interactions are described by the non-local self-energy, we do not require effective interactions $U$ in the strongly correlated orbitals. 
These advantages show that SEET is a promising method to describe physical and chemical properties of challenging strongly correlated, large molecules.

{\it Acknowledgments.}  D.Z.  acknowledges the DOE grant No. ER16391.


\begin{thebibliography}{47}%
\makeatletter
\providecommand \@ifxundefined [1]{%
 \@ifx{#1\undefined}
}%
\providecommand \@ifnum [1]{%
 \ifnum #1\expandafter \@firstoftwo
 \else \expandafter \@secondoftwo
 \fi
}%
\providecommand \@ifx [1]{%
 \ifx #1\expandafter \@firstoftwo
 \else \expandafter \@secondoftwo
 \fi
}%
\providecommand \natexlab [1]{#1}%
\providecommand \enquote  [1]{``#1''}%
\providecommand \bibnamefont  [1]{#1}%
\providecommand \bibfnamefont [1]{#1}%
\providecommand \citenamefont [1]{#1}%
\providecommand \href@noop [0]{\@secondoftwo}%
\providecommand \href [0]{\begingroup \@sanitize@url \@href}%
\providecommand \@href[1]{\@@startlink{#1}\@@href}%
\providecommand \@@href[1]{\endgroup#1\@@endlink}%
\providecommand \@sanitize@url [0]{\catcode `\\12\catcode `\$12\catcode
  `\&12\catcode `\#12\catcode `\^12\catcode `\_12\catcode `\%12\relax}%
\providecommand \@@startlink[1]{}%
\providecommand \@@endlink[0]{}%
\providecommand \url  [0]{\begingroup\@sanitize@url \@url }%
\providecommand \@url [1]{\endgroup\@href {#1}{\urlprefix }}%
\providecommand \urlprefix  [0]{URL }%
\providecommand \Eprint [0]{\href }%
\providecommand \doibase [0]{http://dx.doi.org/}%
\providecommand \selectlanguage [0]{\@gobble}%
\providecommand \bibinfo  [0]{\@secondoftwo}%
\providecommand \bibfield  [0]{\@secondoftwo}%
\providecommand \translation [1]{[#1]}%
\providecommand \BibitemOpen [0]{}%
\providecommand \bibitemStop [0]{}%
\providecommand \bibitemNoStop [0]{.\EOS\space}%
\providecommand \EOS [0]{\spacefactor3000\relax}%
\providecommand \BibitemShut  [1]{\csname bibitem#1\endcsname}%
\let\auto@bib@innerbib\@empty
\bibitem [{\citenamefont {Andersson}\ \emph {et~al.}(1990)\citenamefont
  {Andersson}, \citenamefont {Malmqvist}, \citenamefont {Roos}, \citenamefont
  {Sadlej},\ and\ \citenamefont {Wolinski}}]{andersson1990second}%
  \BibitemOpen
  \bibfield  {author} {\bibinfo {author} {\bibfnamefont {K.}~\bibnamefont
  {Andersson}}, \bibinfo {author} {\bibfnamefont {P.~A.}\ \bibnamefont
  {Malmqvist}}, \bibinfo {author} {\bibfnamefont {B.~O.}\ \bibnamefont {Roos}},
  \bibinfo {author} {\bibfnamefont {A.~J.}\ \bibnamefont {Sadlej}}, \ and\
  \bibinfo {author} {\bibfnamefont {K.}~\bibnamefont {Wolinski}},\ }\href@noop
  {} {\bibfield  {journal} {\bibinfo  {journal} {Journal of Physical
  Chemistry}\ }\textbf {\bibinfo {volume} {94}},\ \bibinfo {pages} {5483}
  (\bibinfo {year} {1990})}\BibitemShut {NoStop}%
\bibitem [{\citenamefont {Ghigo}\ \emph {et~al.}(2004)\citenamefont {Ghigo},
  \citenamefont {Roos},\ and\ \citenamefont {Malmqvist}}]{ghigo2004modified}%
  \BibitemOpen
  \bibfield  {author} {\bibinfo {author} {\bibfnamefont {G.}~\bibnamefont
  {Ghigo}}, \bibinfo {author} {\bibfnamefont {B.~O.}\ \bibnamefont {Roos}}, \
  and\ \bibinfo {author} {\bibfnamefont {P.~A.}\ \bibnamefont {Malmqvist}},\
  }\href@noop {} {\bibfield  {journal} {\bibinfo  {journal} {Chemical physics
  letters}\ }\textbf {\bibinfo {volume} {396}},\ \bibinfo {pages} {142}
  (\bibinfo {year} {2004})}\BibitemShut {NoStop}%
\bibitem [{\citenamefont {Angeli}\ \emph {et~al.}(2001)\citenamefont {Angeli},
  \citenamefont {Cimiraglia}, \citenamefont {Evangelisti}, \citenamefont
  {Leininger},\ and\ \citenamefont {Malrieu}}]{angeli2001introduction}%
  \BibitemOpen
  \bibfield  {author} {\bibinfo {author} {\bibfnamefont {C.}~\bibnamefont
  {Angeli}}, \bibinfo {author} {\bibfnamefont {R.}~\bibnamefont {Cimiraglia}},
  \bibinfo {author} {\bibfnamefont {S.}~\bibnamefont {Evangelisti}}, \bibinfo
  {author} {\bibfnamefont {T.}~\bibnamefont {Leininger}}, \ and\ \bibinfo
  {author} {\bibfnamefont {J.-P.}\ \bibnamefont {Malrieu}},\ }\href@noop {}
  {\bibfield  {journal} {\bibinfo  {journal} {The Journal of Chemical Physics}\
  }\textbf {\bibinfo {volume} {114}},\ \bibinfo {pages} {10252} (\bibinfo
  {year} {2001})}\BibitemShut {NoStop}%
\bibitem [{\citenamefont {Angeli}\ \emph {et~al.}(2002)\citenamefont {Angeli},
  \citenamefont {Cimiraglia},\ and\ \citenamefont {Malrieu}}]{angeli2002n}%
  \BibitemOpen
  \bibfield  {author} {\bibinfo {author} {\bibfnamefont {C.}~\bibnamefont
  {Angeli}}, \bibinfo {author} {\bibfnamefont {R.}~\bibnamefont {Cimiraglia}},
  \ and\ \bibinfo {author} {\bibfnamefont {J.-P.}\ \bibnamefont {Malrieu}},\
  }\href@noop {} {\bibfield  {journal} {\bibinfo  {journal} {The Journal of
  chemical physics}\ }\textbf {\bibinfo {volume} {117}},\ \bibinfo {pages}
  {9138} (\bibinfo {year} {2002})}\BibitemShut {NoStop}%
\bibitem [{\citenamefont {White}\ and\ \citenamefont
  {Martin}(1999)}]{white1999ab}%
  \BibitemOpen
  \bibfield  {author} {\bibinfo {author} {\bibfnamefont {S.~R.}\ \bibnamefont
  {White}}\ and\ \bibinfo {author} {\bibfnamefont {R.~L.}\ \bibnamefont
  {Martin}},\ }\href@noop {} {\bibfield  {journal} {\bibinfo  {journal} {The
  Journal of chemical physics}\ }\textbf {\bibinfo {volume} {110}},\ \bibinfo
  {pages} {4127} (\bibinfo {year} {1999})}\BibitemShut {NoStop}%
\bibitem [{\citenamefont {Chan}\ and\ \citenamefont
  {Zgid}(2009)}]{chan2009density}%
  \BibitemOpen
  \bibfield  {author} {\bibinfo {author} {\bibfnamefont {G.~K.-L.}\
  \bibnamefont {Chan}}\ and\ \bibinfo {author} {\bibfnamefont {D.}~\bibnamefont
  {Zgid}},\ }\href@noop {} {\bibfield  {journal} {\bibinfo  {journal} {Annual
  Reports in Computational Chemistry}\ }\textbf {\bibinfo {volume} {5}},\
  \bibinfo {pages} {149} (\bibinfo {year} {2009})}\BibitemShut {NoStop}%
\bibitem [{\citenamefont {Zgid}\ and\ \citenamefont
  {Nooijen}(2008)}]{zgid2008density}%
  \BibitemOpen
  \bibfield  {author} {\bibinfo {author} {\bibfnamefont {D.}~\bibnamefont
  {Zgid}}\ and\ \bibinfo {author} {\bibfnamefont {M.}~\bibnamefont {Nooijen}},\
  }\href@noop {} {\bibfield  {journal} {\bibinfo  {journal} {The Journal of
  chemical physics}\ }\textbf {\bibinfo {volume} {128}},\ \bibinfo {pages}
  {144116} (\bibinfo {year} {2008})}\BibitemShut {NoStop}%
\bibitem [{\citenamefont {Kurashige}\ and\ \citenamefont
  {Yanai}(2009)}]{kurashige2009high}%
  \BibitemOpen
  \bibfield  {author} {\bibinfo {author} {\bibfnamefont {Y.}~\bibnamefont
  {Kurashige}}\ and\ \bibinfo {author} {\bibfnamefont {T.}~\bibnamefont
  {Yanai}},\ }\href@noop {} {\bibfield  {journal} {\bibinfo  {journal} {The
  Journal of chemical physics}\ }\textbf {\bibinfo {volume} {130}},\ \bibinfo
  {pages} {234114} (\bibinfo {year} {2009})}\BibitemShut {NoStop}%
\bibitem [{\citenamefont {Marti}\ and\ \citenamefont
  {Reiher}(2011)}]{marti2011new}%
  \BibitemOpen
  \bibfield  {author} {\bibinfo {author} {\bibfnamefont {K.~H.}\ \bibnamefont
  {Marti}}\ and\ \bibinfo {author} {\bibfnamefont {M.}~\bibnamefont {Reiher}},\
  }\href@noop {} {\bibfield  {journal} {\bibinfo  {journal} {Physical Chemistry
  Chemical Physics}\ }\textbf {\bibinfo {volume} {13}},\ \bibinfo {pages}
  {6750} (\bibinfo {year} {2011})}\BibitemShut {NoStop}%
\bibitem [{\citenamefont {Malmqvist}\ \emph {et~al.}(1990)\citenamefont
  {Malmqvist}, \citenamefont {Rendell},\ and\ \citenamefont
  {Roos}}]{malmqvist1990restricted}%
  \BibitemOpen
  \bibfield  {author} {\bibinfo {author} {\bibfnamefont {P.~A.}\ \bibnamefont
  {Malmqvist}}, \bibinfo {author} {\bibfnamefont {A.}~\bibnamefont {Rendell}},
  \ and\ \bibinfo {author} {\bibfnamefont {B.~O.}\ \bibnamefont {Roos}},\
  }\href@noop {} {\bibfield  {journal} {\bibinfo  {journal} {Journal of
  Physical Chemistry}\ }\textbf {\bibinfo {volume} {94}},\ \bibinfo {pages}
  {5477} (\bibinfo {year} {1990})}\BibitemShut {NoStop}%
\bibitem [{\citenamefont {Malmqvist}\ \emph {et~al.}(2008)\citenamefont
  {Malmqvist}, \citenamefont {Pierloot}, \citenamefont {Shahi}, \citenamefont
  {Cramer},\ and\ \citenamefont {Gagliardi}}]{malmqvist2008restricted}%
  \BibitemOpen
  \bibfield  {author} {\bibinfo {author} {\bibfnamefont {P.~{\AA}.}\
  \bibnamefont {Malmqvist}}, \bibinfo {author} {\bibfnamefont {K.}~\bibnamefont
  {Pierloot}}, \bibinfo {author} {\bibfnamefont {A.~R.~M.}\ \bibnamefont
  {Shahi}}, \bibinfo {author} {\bibfnamefont {C.~J.}\ \bibnamefont {Cramer}}, \
  and\ \bibinfo {author} {\bibfnamefont {L.}~\bibnamefont {Gagliardi}},\
  }\href@noop {} {\bibfield  {journal} {\bibinfo  {journal} {The Journal of
  chemical physics}\ }\textbf {\bibinfo {volume} {128}},\ \bibinfo {pages}
  {204109} (\bibinfo {year} {2008})}\BibitemShut {NoStop}%
\bibitem [{\citenamefont {Manni}\ \emph {et~al.}(2011)\citenamefont {Manni},
  \citenamefont {Aquilante},\ and\ \citenamefont
  {Gagliardi}}]{manni2011strong}%
  \BibitemOpen
  \bibfield  {author} {\bibinfo {author} {\bibfnamefont {G.~L.}\ \bibnamefont
  {Manni}}, \bibinfo {author} {\bibfnamefont {F.}~\bibnamefont {Aquilante}}, \
  and\ \bibinfo {author} {\bibfnamefont {L.}~\bibnamefont {Gagliardi}},\
  }\href@noop {} {\bibfield  {journal} {\bibinfo  {journal} {The Journal of
  chemical physics}\ }\textbf {\bibinfo {volume} {134}},\ \bibinfo {pages}
  {034114} (\bibinfo {year} {2011})}\BibitemShut {NoStop}%
\bibitem [{\citenamefont {Li~Manni}\ \emph {et~al.}(2013)\citenamefont
  {Li~Manni}, \citenamefont {Ma}, \citenamefont {Aquilante}, \citenamefont
  {Olsen},\ and\ \citenamefont {Gagliardi}}]{li2013splitgas}%
  \BibitemOpen
  \bibfield  {author} {\bibinfo {author} {\bibfnamefont {G.}~\bibnamefont
  {Li~Manni}}, \bibinfo {author} {\bibfnamefont {D.}~\bibnamefont {Ma}},
  \bibinfo {author} {\bibfnamefont {F.}~\bibnamefont {Aquilante}}, \bibinfo
  {author} {\bibfnamefont {J.}~\bibnamefont {Olsen}}, \ and\ \bibinfo {author}
  {\bibfnamefont {L.}~\bibnamefont {Gagliardi}},\ }\href@noop {} {\bibfield
  {journal} {\bibinfo  {journal} {Journal of Chemical Theory and Computation}\
  }\textbf {\bibinfo {volume} {9}},\ \bibinfo {pages} {3375} (\bibinfo {year}
  {2013})}\BibitemShut {NoStop}%
\bibitem [{\citenamefont {Ma}\ \emph {et~al.}(2011)\citenamefont {Ma},
  \citenamefont {Manni},\ and\ \citenamefont {Gagliardi}}]{ma2011generalized}%
  \BibitemOpen
  \bibfield  {author} {\bibinfo {author} {\bibfnamefont {D.}~\bibnamefont
  {Ma}}, \bibinfo {author} {\bibfnamefont {G.~L.}\ \bibnamefont {Manni}}, \
  and\ \bibinfo {author} {\bibfnamefont {L.}~\bibnamefont {Gagliardi}},\
  }\href@noop {} {\bibfield  {journal} {\bibinfo  {journal} {The Journal of
  chemical physics}\ }\textbf {\bibinfo {volume} {135}},\ \bibinfo {pages}
  {044128} (\bibinfo {year} {2011})}\BibitemShut {NoStop}%
\bibitem [{\citenamefont {Vogiatzis}\ \emph {et~al.}(2015)\citenamefont
  {Vogiatzis}, \citenamefont {Li~Manni}, \citenamefont {Stoneburner},
  \citenamefont {Ma},\ and\ \citenamefont
  {Gagliardi}}]{vogiatzis2015systematic}%
  \BibitemOpen
  \bibfield  {author} {\bibinfo {author} {\bibfnamefont {K.~D.}\ \bibnamefont
  {Vogiatzis}}, \bibinfo {author} {\bibfnamefont {G.}~\bibnamefont {Li~Manni}},
  \bibinfo {author} {\bibfnamefont {S.~J.}\ \bibnamefont {Stoneburner}},
  \bibinfo {author} {\bibfnamefont {D.}~\bibnamefont {Ma}}, \ and\ \bibinfo
  {author} {\bibfnamefont {L.}~\bibnamefont {Gagliardi}},\ }\href@noop {}
  {\bibfield  {journal} {\bibinfo  {journal} {Journal of Chemical Theory and
  Computation}\ } (\bibinfo {year} {2015})}\BibitemShut {NoStop}%
\bibitem [{\citenamefont {Gr{\"a}fenstein}\ and\ \citenamefont
  {Cremer}(2000)}]{grafenstein2000combination}%
  \BibitemOpen
  \bibfield  {author} {\bibinfo {author} {\bibfnamefont {J.}~\bibnamefont
  {Gr{\"a}fenstein}}\ and\ \bibinfo {author} {\bibfnamefont {D.}~\bibnamefont
  {Cremer}},\ }\href@noop {} {\bibfield  {journal} {\bibinfo  {journal}
  {Chemical Physics Letters}\ }\textbf {\bibinfo {volume} {316}},\ \bibinfo
  {pages} {569} (\bibinfo {year} {2000})}\BibitemShut {NoStop}%
\bibitem [{\citenamefont {Gr{\"a}fenstein}\ and\ \citenamefont
  {Cremer}(2005)}]{grafenstein2005development}%
  \BibitemOpen
  \bibfield  {author} {\bibinfo {author} {\bibfnamefont {J.}~\bibnamefont
  {Gr{\"a}fenstein}}\ and\ \bibinfo {author} {\bibfnamefont {D.}~\bibnamefont
  {Cremer}},\ }\href@noop {} {\bibfield  {journal} {\bibinfo  {journal}
  {Molecular Physics}\ }\textbf {\bibinfo {volume} {103}},\ \bibinfo {pages}
  {279} (\bibinfo {year} {2005})}\BibitemShut {NoStop}%
\bibitem [{\citenamefont {Pollet}\ \emph {et~al.}(2002)\citenamefont {Pollet},
  \citenamefont {Savin}, \citenamefont {Leininger},\ and\ \citenamefont
  {Stoll}}]{savin_mc_dft_2002}%
  \BibitemOpen
  \bibfield  {author} {\bibinfo {author} {\bibfnamefont {R.}~\bibnamefont
  {Pollet}}, \bibinfo {author} {\bibfnamefont {A.}~\bibnamefont {Savin}},
  \bibinfo {author} {\bibfnamefont {T.}~\bibnamefont {Leininger}}, \ and\
  \bibinfo {author} {\bibfnamefont {H.}~\bibnamefont {Stoll}},\ }\href
  {\doibase http://dx.doi.org/10.1063/1.1430739} {\bibfield  {journal}
  {\bibinfo  {journal} {The Journal of Chemical Physics}\ }\textbf {\bibinfo
  {volume} {116}},\ \bibinfo {pages} {1250} (\bibinfo {year}
  {2002})}\BibitemShut {NoStop}%
\bibitem [{\citenamefont {Li~Manni}\ \emph {et~al.}(2014)\citenamefont
  {Li~Manni}, \citenamefont {Carlson}, \citenamefont {Luo}, \citenamefont {Ma},
  \citenamefont {Olsen}, \citenamefont {Truhlar},\ and\ \citenamefont
  {Gagliardi}}]{li2014multiconfiguration}%
  \BibitemOpen
  \bibfield  {author} {\bibinfo {author} {\bibfnamefont {G.}~\bibnamefont
  {Li~Manni}}, \bibinfo {author} {\bibfnamefont {R.~K.}\ \bibnamefont
  {Carlson}}, \bibinfo {author} {\bibfnamefont {S.}~\bibnamefont {Luo}},
  \bibinfo {author} {\bibfnamefont {D.}~\bibnamefont {Ma}}, \bibinfo {author}
  {\bibfnamefont {J.}~\bibnamefont {Olsen}}, \bibinfo {author} {\bibfnamefont
  {D.~G.}\ \bibnamefont {Truhlar}}, \ and\ \bibinfo {author} {\bibfnamefont
  {L.}~\bibnamefont {Gagliardi}},\ }\href@noop {} {\bibfield  {journal}
  {\bibinfo  {journal} {Journal of Chemical Theory and Computation}\ }\textbf
  {\bibinfo {volume} {10}},\ \bibinfo {pages} {3669} (\bibinfo {year}
  {2014})}\BibitemShut {NoStop}%
\bibitem [{\citenamefont {Carlson}\ \emph {et~al.}(2015)\citenamefont
  {Carlson}, \citenamefont {Truhlar},\ and\ \citenamefont
  {Gagliardi}}]{carlson2015multiconfiguration}%
  \BibitemOpen
  \bibfield  {author} {\bibinfo {author} {\bibfnamefont {R.~K.}\ \bibnamefont
  {Carlson}}, \bibinfo {author} {\bibfnamefont {D.~G.}\ \bibnamefont
  {Truhlar}}, \ and\ \bibinfo {author} {\bibfnamefont {L.}~\bibnamefont
  {Gagliardi}},\ }\href@noop {} {\bibfield  {journal} {\bibinfo  {journal}
  {Journal of Chemical Theory and Computation}\ }\textbf {\bibinfo {volume}
  {11}},\ \bibinfo {pages} {4077} (\bibinfo {year} {2015})}\BibitemShut
  {NoStop}%
\bibitem [{\citenamefont {Georges}\ \emph {et~al.}(1996)\citenamefont
  {Georges}, \citenamefont {Kotliar}, \citenamefont {Krauth},\ and\
  \citenamefont {Rozenberg}}]{georges1996dynamical}%
  \BibitemOpen
  \bibfield  {author} {\bibinfo {author} {\bibfnamefont {A.}~\bibnamefont
  {Georges}}, \bibinfo {author} {\bibfnamefont {G.}~\bibnamefont {Kotliar}},
  \bibinfo {author} {\bibfnamefont {W.}~\bibnamefont {Krauth}}, \ and\ \bibinfo
  {author} {\bibfnamefont {M.~J.}\ \bibnamefont {Rozenberg}},\ }\href@noop {}
  {\bibfield  {journal} {\bibinfo  {journal} {Reviews of Modern Physics}\
  }\textbf {\bibinfo {volume} {68}},\ \bibinfo {pages} {13} (\bibinfo {year}
  {1996})}\BibitemShut {NoStop}%
\bibitem [{\citenamefont {Kotliar}\ \emph {et~al.}(2006)\citenamefont
  {Kotliar}, \citenamefont {Savrasov}, \citenamefont {Haule}, \citenamefont
  {Oudovenko}, \citenamefont {Parcollet},\ and\ \citenamefont
  {Marianetti}}]{kotliar2006electronic}%
  \BibitemOpen
  \bibfield  {author} {\bibinfo {author} {\bibfnamefont {G.}~\bibnamefont
  {Kotliar}}, \bibinfo {author} {\bibfnamefont {S.~Y.}\ \bibnamefont
  {Savrasov}}, \bibinfo {author} {\bibfnamefont {K.}~\bibnamefont {Haule}},
  \bibinfo {author} {\bibfnamefont {V.~S.}\ \bibnamefont {Oudovenko}}, \bibinfo
  {author} {\bibfnamefont {O.}~\bibnamefont {Parcollet}}, \ and\ \bibinfo
  {author} {\bibfnamefont {C.}~\bibnamefont {Marianetti}},\ }\href@noop {}
  {\bibfield  {journal} {\bibinfo  {journal} {Reviews of Modern Physics}\
  }\textbf {\bibinfo {volume} {78}},\ \bibinfo {pages} {865} (\bibinfo {year}
  {2006})}\BibitemShut {NoStop}%
\bibitem [{\citenamefont {Kotliar}\ and\ \citenamefont
  {Vollhardt}(2004)}]{kotliar2004strongly}%
  \BibitemOpen
  \bibfield  {author} {\bibinfo {author} {\bibfnamefont {G.}~\bibnamefont
  {Kotliar}}\ and\ \bibinfo {author} {\bibfnamefont {D.}~\bibnamefont
  {Vollhardt}},\ }\href@noop {} {\bibfield  {journal} {\bibinfo  {journal}
  {Physics Today}\ }\textbf {\bibinfo {volume} {57}},\ \bibinfo {pages} {53}
  (\bibinfo {year} {2004})}\BibitemShut {NoStop}%
\bibitem [{\citenamefont {Held}(2007)}]{held2007electronic}%
  \BibitemOpen
  \bibfield  {author} {\bibinfo {author} {\bibfnamefont {K.}~\bibnamefont
  {Held}},\ }\href@noop {} {\bibfield  {journal} {\bibinfo  {journal} {Advances
  in Physics}\ }\textbf {\bibinfo {volume} {56}},\ \bibinfo {pages} {829}
  (\bibinfo {year} {2007})}\BibitemShut {NoStop}%
\bibitem [{\citenamefont {Lechermann}\ \emph {et~al.}(2006)\citenamefont
  {Lechermann}, \citenamefont {Georges}, \citenamefont {Poteryaev},
  \citenamefont {Biermann}, \citenamefont {Posternak}, \citenamefont
  {Yamasaki},\ and\ \citenamefont {Andersen}}]{lechermann2006dynamical}%
  \BibitemOpen
  \bibfield  {author} {\bibinfo {author} {\bibfnamefont {F.}~\bibnamefont
  {Lechermann}}, \bibinfo {author} {\bibfnamefont {A.}~\bibnamefont {Georges}},
  \bibinfo {author} {\bibfnamefont {A.}~\bibnamefont {Poteryaev}}, \bibinfo
  {author} {\bibfnamefont {S.}~\bibnamefont {Biermann}}, \bibinfo {author}
  {\bibfnamefont {M.}~\bibnamefont {Posternak}}, \bibinfo {author}
  {\bibfnamefont {A.}~\bibnamefont {Yamasaki}}, \ and\ \bibinfo {author}
  {\bibfnamefont {O.}~\bibnamefont {Andersen}},\ }\href@noop {} {\bibfield
  {journal} {\bibinfo  {journal} {Physical Review B}\ }\textbf {\bibinfo
  {volume} {74}},\ \bibinfo {pages} {125120} (\bibinfo {year}
  {2006})}\BibitemShut {NoStop}%
\bibitem [{\citenamefont {Lichtenstein}\ \emph {et~al.}(2001)\citenamefont
  {Lichtenstein}, \citenamefont {Katsnelson},\ and\ \citenamefont
  {Kotliar}}]{lichtenstein2001finite}%
  \BibitemOpen
  \bibfield  {author} {\bibinfo {author} {\bibfnamefont {A.~I.}\ \bibnamefont
  {Lichtenstein}}, \bibinfo {author} {\bibfnamefont {M.~I.}\ \bibnamefont
  {Katsnelson}}, \ and\ \bibinfo {author} {\bibfnamefont {G.}~\bibnamefont
  {Kotliar}},\ }\href {\doibase 10.1103/PhysRevLett.87.067205} {\bibfield
  {journal} {\bibinfo  {journal} {Phys. Rev. Lett.}\ }\textbf {\bibinfo
  {volume} {87}},\ \bibinfo {pages} {067205} (\bibinfo {year}
  {2001})}\BibitemShut {NoStop}%
\bibitem [{\citenamefont {Ren}\ \emph {et~al.}(2006)\citenamefont {Ren},
  \citenamefont {Leonov}, \citenamefont {Keller}, \citenamefont {Kollar},
  \citenamefont {Nekrasov},\ and\ \citenamefont {Vollhardt}}]{ren2006lda+}%
  \BibitemOpen
  \bibfield  {author} {\bibinfo {author} {\bibfnamefont {X.}~\bibnamefont
  {Ren}}, \bibinfo {author} {\bibfnamefont {I.}~\bibnamefont {Leonov}},
  \bibinfo {author} {\bibfnamefont {G.}~\bibnamefont {Keller}}, \bibinfo
  {author} {\bibfnamefont {M.}~\bibnamefont {Kollar}}, \bibinfo {author}
  {\bibfnamefont {I.}~\bibnamefont {Nekrasov}}, \ and\ \bibinfo {author}
  {\bibfnamefont {D.}~\bibnamefont {Vollhardt}},\ }\href@noop {} {\bibfield
  {journal} {\bibinfo  {journal} {Physical Review B}\ }\textbf {\bibinfo
  {volume} {74}},\ \bibinfo {pages} {195114} (\bibinfo {year}
  {2006})}\BibitemShut {NoStop}%
\bibitem [{\citenamefont {Lee}\ and\ \citenamefont
  {Haule}(2015)}]{lee2015dynamical}%
  \BibitemOpen
  \bibfield  {author} {\bibinfo {author} {\bibfnamefont {J.}~\bibnamefont
  {Lee}}\ and\ \bibinfo {author} {\bibfnamefont {K.}~\bibnamefont {Haule}},\
  }\href@noop {} {\bibfield  {journal} {\bibinfo  {journal} {Physical Review
  B}\ }\textbf {\bibinfo {volume} {91}},\ \bibinfo {pages} {155144} (\bibinfo
  {year} {2015})}\BibitemShut {NoStop}%
\bibitem [{\citenamefont {Boukhvalov}\ \emph {et~al.}(2008)\citenamefont
  {Boukhvalov}, \citenamefont {Vergara}, \citenamefont {Dobrovitski},
  \citenamefont {Katsnelson}, \citenamefont {Lichtenstein}, \citenamefont
  {K{\"o}gerler}, \citenamefont {Musfeldt},\ and\ \citenamefont
  {Harmon}}]{boukhvalov2008correlation}%
  \BibitemOpen
  \bibfield  {author} {\bibinfo {author} {\bibfnamefont {D.}~\bibnamefont
  {Boukhvalov}}, \bibinfo {author} {\bibfnamefont {L.}~\bibnamefont {Vergara}},
  \bibinfo {author} {\bibfnamefont {V.}~\bibnamefont {Dobrovitski}}, \bibinfo
  {author} {\bibfnamefont {M.}~\bibnamefont {Katsnelson}}, \bibinfo {author}
  {\bibfnamefont {A.}~\bibnamefont {Lichtenstein}}, \bibinfo {author}
  {\bibfnamefont {P.}~\bibnamefont {K{\"o}gerler}}, \bibinfo {author}
  {\bibfnamefont {J.}~\bibnamefont {Musfeldt}}, \ and\ \bibinfo {author}
  {\bibfnamefont {B.}~\bibnamefont {Harmon}},\ }\href@noop {} {\bibfield
  {journal} {\bibinfo  {journal} {Physical Review B}\ }\textbf {\bibinfo
  {volume} {77}},\ \bibinfo {pages} {180402} (\bibinfo {year}
  {2008})}\BibitemShut {NoStop}%
\bibitem [{\citenamefont {Weber}\ \emph {et~al.}(2014)\citenamefont {Weber},
  \citenamefont {Cole}, \citenamefont {O'Regan},\ and\ \citenamefont
  {Payne}}]{weber2014renormalization}%
  \BibitemOpen
  \bibfield  {author} {\bibinfo {author} {\bibfnamefont {C.}~\bibnamefont
  {Weber}}, \bibinfo {author} {\bibfnamefont {D.~J.}\ \bibnamefont {Cole}},
  \bibinfo {author} {\bibfnamefont {D.~D.}\ \bibnamefont {O'Regan}}, \ and\
  \bibinfo {author} {\bibfnamefont {M.~C.}\ \bibnamefont {Payne}},\ }\href@noop
  {} {\bibfield  {journal} {\bibinfo  {journal} {Proceedings of the National
  Academy of Sciences}\ }\textbf {\bibinfo {volume} {111}},\ \bibinfo {pages}
  {5790} (\bibinfo {year} {2014})}\BibitemShut {NoStop}%
\bibitem [{\citenamefont {Weber}\ \emph {et~al.}(2013)\citenamefont {Weber},
  \citenamefont {O'Regan}, \citenamefont {Hine}, \citenamefont {Littlewood},
  \citenamefont {Kotliar},\ and\ \citenamefont {Payne}}]{weber2013importance}%
  \BibitemOpen
  \bibfield  {author} {\bibinfo {author} {\bibfnamefont {C.}~\bibnamefont
  {Weber}}, \bibinfo {author} {\bibfnamefont {D.~D.}\ \bibnamefont {O'Regan}},
  \bibinfo {author} {\bibfnamefont {N.~D.}\ \bibnamefont {Hine}}, \bibinfo
  {author} {\bibfnamefont {P.~B.}\ \bibnamefont {Littlewood}}, \bibinfo
  {author} {\bibfnamefont {G.}~\bibnamefont {Kotliar}}, \ and\ \bibinfo
  {author} {\bibfnamefont {M.~C.}\ \bibnamefont {Payne}},\ }\href@noop {}
  {\bibfield  {journal} {\bibinfo  {journal} {Physical review letters}\
  }\textbf {\bibinfo {volume} {110}},\ \bibinfo {pages} {106402} (\bibinfo
  {year} {2013})}\BibitemShut {NoStop}%
\bibitem [{\citenamefont {Chen}\ \emph {et~al.}(2015)\citenamefont {Chen},
  \citenamefont {Millis},\ and\ \citenamefont
  {Marianetti}}]{PhysRevB.91.241111}%
  \BibitemOpen
  \bibfield  {author} {\bibinfo {author} {\bibfnamefont {J.}~\bibnamefont
  {Chen}}, \bibinfo {author} {\bibfnamefont {A.~J.}\ \bibnamefont {Millis}}, \
  and\ \bibinfo {author} {\bibfnamefont {C.~A.}\ \bibnamefont {Marianetti}},\
  }\href {\doibase 10.1103/PhysRevB.91.241111} {\bibfield  {journal} {\bibinfo
  {journal} {Phys. Rev. B}\ }\textbf {\bibinfo {volume} {91}},\ \bibinfo
  {pages} {241111} (\bibinfo {year} {2015})}\BibitemShut {NoStop}%
\bibitem [{\citenamefont {Park}\ \emph {et~al.}(2015)\citenamefont {Park},
  \citenamefont {Millis},\ and\ \citenamefont
  {Marianetti}}]{PhysRevB.92.035146}%
  \BibitemOpen
  \bibfield  {author} {\bibinfo {author} {\bibfnamefont {H.}~\bibnamefont
  {Park}}, \bibinfo {author} {\bibfnamefont {A.~J.}\ \bibnamefont {Millis}}, \
  and\ \bibinfo {author} {\bibfnamefont {C.~A.}\ \bibnamefont {Marianetti}},\
  }\href {\doibase 10.1103/PhysRevB.92.035146} {\bibfield  {journal} {\bibinfo
  {journal} {Phys. Rev. B}\ }\textbf {\bibinfo {volume} {92}},\ \bibinfo
  {pages} {035146} (\bibinfo {year} {2015})}\BibitemShut {NoStop}%
\bibitem [{\citenamefont {Vaugier}\ \emph {et~al.}(2012)\citenamefont
  {Vaugier}, \citenamefont {Jiang},\ and\ \citenamefont
  {Biermann}}]{vaugier2012hubbard}%
  \BibitemOpen
  \bibfield  {author} {\bibinfo {author} {\bibfnamefont {L.}~\bibnamefont
  {Vaugier}}, \bibinfo {author} {\bibfnamefont {H.}~\bibnamefont {Jiang}}, \
  and\ \bibinfo {author} {\bibfnamefont {S.}~\bibnamefont {Biermann}},\
  }\href@noop {} {\bibfield  {journal} {\bibinfo  {journal} {Physical Review
  B}\ }\textbf {\bibinfo {volume} {86}},\ \bibinfo {pages} {165105} (\bibinfo
  {year} {2012})}\BibitemShut {NoStop}%
\bibitem [{\citenamefont {Biermann}\ \emph {et~al.}(2003)\citenamefont
  {Biermann}, \citenamefont {Aryasetiawan},\ and\ \citenamefont
  {Georges}}]{biermann2003first}%
  \BibitemOpen
  \bibfield  {author} {\bibinfo {author} {\bibfnamefont {S.}~\bibnamefont
  {Biermann}}, \bibinfo {author} {\bibfnamefont {F.}~\bibnamefont
  {Aryasetiawan}}, \ and\ \bibinfo {author} {\bibfnamefont {A.}~\bibnamefont
  {Georges}},\ }\href@noop {} {\bibfield  {journal} {\bibinfo  {journal}
  {Physical review letters}\ }\textbf {\bibinfo {volume} {90}},\ \bibinfo
  {pages} {086402} (\bibinfo {year} {2003})}\BibitemShut {NoStop}%
\bibitem [{\citenamefont {Tomczak}\ \emph {et~al.}(2012)\citenamefont
  {Tomczak}, \citenamefont {Casula}, \citenamefont {Miyake}, \citenamefont
  {Aryasetiawan},\ and\ \citenamefont {Biermann}}]{tomczak2012combined}%
  \BibitemOpen
  \bibfield  {author} {\bibinfo {author} {\bibfnamefont {J.~M.}\ \bibnamefont
  {Tomczak}}, \bibinfo {author} {\bibfnamefont {M.}~\bibnamefont {Casula}},
  \bibinfo {author} {\bibfnamefont {T.}~\bibnamefont {Miyake}}, \bibinfo
  {author} {\bibfnamefont {F.}~\bibnamefont {Aryasetiawan}}, \ and\ \bibinfo
  {author} {\bibfnamefont {S.}~\bibnamefont {Biermann}},\ }\href@noop {}
  {\bibfield  {journal} {\bibinfo  {journal} {EPL (Europhysics Letters)}\
  }\textbf {\bibinfo {volume} {100}},\ \bibinfo {pages} {67001} (\bibinfo
  {year} {2012})}\BibitemShut {NoStop}%
\bibitem [{\citenamefont {Kitatani}\ \emph {et~al.}(2015)\citenamefont
  {Kitatani}, \citenamefont {Tsuji},\ and\ \citenamefont
  {Aoki}}]{PhysRevB.92.085104}%
  \BibitemOpen
  \bibfield  {author} {\bibinfo {author} {\bibfnamefont {M.}~\bibnamefont
  {Kitatani}}, \bibinfo {author} {\bibfnamefont {N.}~\bibnamefont {Tsuji}}, \
  and\ \bibinfo {author} {\bibfnamefont {H.}~\bibnamefont {Aoki}},\ }\href
  {\doibase 10.1103/PhysRevB.92.085104} {\bibfield  {journal} {\bibinfo
  {journal} {Phys. Rev. B}\ }\textbf {\bibinfo {volume} {92}},\ \bibinfo
  {pages} {085104} (\bibinfo {year} {2015})}\BibitemShut {NoStop}%
\bibitem [{\citenamefont {Kananenka}\ \emph {et~al.}(2015)\citenamefont
  {Kananenka}, \citenamefont {Gull},\ and\ \citenamefont
  {Zgid}}]{kananenka2015systematically}%
  \BibitemOpen
  \bibfield  {author} {\bibinfo {author} {\bibfnamefont {A.~A.}\ \bibnamefont
  {Kananenka}}, \bibinfo {author} {\bibfnamefont {E.}~\bibnamefont {Gull}}, \
  and\ \bibinfo {author} {\bibfnamefont {D.}~\bibnamefont {Zgid}},\ }\href@noop
  {} {\bibfield  {journal} {\bibinfo  {journal} {Physical Review B}\ }\textbf
  {\bibinfo {volume} {91}},\ \bibinfo {pages} {121111} (\bibinfo {year}
  {2015})}\BibitemShut {NoStop}%
\bibitem [{\citenamefont {Phillips}\ and\ \citenamefont
  {Zgid}(2014)}]{phillips2014communication}%
  \BibitemOpen
  \bibfield  {author} {\bibinfo {author} {\bibfnamefont {J.~J.}\ \bibnamefont
  {Phillips}}\ and\ \bibinfo {author} {\bibfnamefont {D.}~\bibnamefont
  {Zgid}},\ }\href@noop {} {\bibfield  {journal} {\bibinfo  {journal} {The
  Journal of chemical physics}\ }\textbf {\bibinfo {volume} {140}},\ \bibinfo
  {pages} {241101} (\bibinfo {year} {2014})}\BibitemShut {NoStop}%
\bibitem [{\citenamefont {Phillips}\ \emph {et~al.}(2015)\citenamefont
  {Phillips}, \citenamefont {Kananenka},\ and\ \citenamefont
  {Zgid}}]{phillips2015fractional}%
  \BibitemOpen
  \bibfield  {author} {\bibinfo {author} {\bibfnamefont {J.~J.}\ \bibnamefont
  {Phillips}}, \bibinfo {author} {\bibfnamefont {A.~A.}\ \bibnamefont
  {Kananenka}}, \ and\ \bibinfo {author} {\bibfnamefont {D.}~\bibnamefont
  {Zgid}},\ }\href@noop {} {\bibfield  {journal} {\bibinfo  {journal} {The
  Journal of chemical physics}\ }\textbf {\bibinfo {volume} {142}},\ \bibinfo
  {pages} {194108} (\bibinfo {year} {2015})}\BibitemShut {NoStop}%
\bibitem [{\citenamefont {Zgid}\ and\ \citenamefont
  {Chan}(2011)}]{zgid2011dynamical}%
  \BibitemOpen
  \bibfield  {author} {\bibinfo {author} {\bibfnamefont {D.}~\bibnamefont
  {Zgid}}\ and\ \bibinfo {author} {\bibfnamefont {G.~K.-L.}\ \bibnamefont
  {Chan}},\ }\href@noop {} {\bibfield  {journal} {\bibinfo  {journal} {The
  Journal of chemical physics}\ }\textbf {\bibinfo {volume} {134}},\ \bibinfo
  {pages} {094115} (\bibinfo {year} {2011})}\BibitemShut {NoStop}%
\bibitem [{\citenamefont {Neese}(2012)}]{neese2012orca}%
  \BibitemOpen
  \bibfield  {author} {\bibinfo {author} {\bibfnamefont {F.}~\bibnamefont
  {Neese}},\ }\href@noop {} {\bibfield  {journal} {\bibinfo  {journal} {Wiley
  Interdisciplinary Reviews: Computational Molecular Science}\ }\textbf
  {\bibinfo {volume} {2}},\ \bibinfo {pages} {73} (\bibinfo {year}
  {2012})}\BibitemShut {NoStop}%
\bibitem [{\citenamefont {Aidas}\ \emph {et~al.}(2014)\citenamefont {Aidas},
  \citenamefont {Angeli}, \citenamefont {Bak}, \citenamefont {Bakken},
  \citenamefont {Bast}, \citenamefont {Boman}, \citenamefont {Christiansen},
  \citenamefont {Cimiraglia}, \citenamefont {Coriani}, \citenamefont {Dahle}
  \emph {et~al.}}]{aidas2014dalton}%
  \BibitemOpen
  \bibfield  {author} {\bibinfo {author} {\bibfnamefont {K.}~\bibnamefont
  {Aidas}}, \bibinfo {author} {\bibfnamefont {C.}~\bibnamefont {Angeli}},
  \bibinfo {author} {\bibfnamefont {K.~L.}\ \bibnamefont {Bak}}, \bibinfo
  {author} {\bibfnamefont {V.}~\bibnamefont {Bakken}}, \bibinfo {author}
  {\bibfnamefont {R.}~\bibnamefont {Bast}}, \bibinfo {author} {\bibfnamefont
  {L.}~\bibnamefont {Boman}}, \bibinfo {author} {\bibfnamefont
  {O.}~\bibnamefont {Christiansen}}, \bibinfo {author} {\bibfnamefont
  {R.}~\bibnamefont {Cimiraglia}}, \bibinfo {author} {\bibfnamefont
  {S.}~\bibnamefont {Coriani}}, \bibinfo {author} {\bibfnamefont
  {P.}~\bibnamefont {Dahle}},  \emph {et~al.},\ }\href@noop {} {\bibfield
  {journal} {\bibinfo  {journal} {Wiley Interdisciplinary Reviews:
  Computational Molecular Science}\ }\textbf {\bibinfo {volume} {4}},\ \bibinfo
  {pages} {269} (\bibinfo {year} {2014})}\BibitemShut {NoStop}%
\bibitem [{\citenamefont {Zgid}\ \emph {et~al.}(2012)\citenamefont {Zgid},
  \citenamefont {Gull},\ and\ \citenamefont {Chan}}]{zgid_chan_solver_prb}%
  \BibitemOpen
  \bibfield  {author} {\bibinfo {author} {\bibfnamefont {D.}~\bibnamefont
  {Zgid}}, \bibinfo {author} {\bibfnamefont {E.}~\bibnamefont {Gull}}, \ and\
  \bibinfo {author} {\bibfnamefont {G.~K.-L.}\ \bibnamefont {Chan}},\ }\href
  {\doibase 10.1103/PhysRevB.86.165128} {\bibfield  {journal} {\bibinfo
  {journal} {Physical Review B}\ }\textbf {\bibinfo {volume} {86}},\ \bibinfo
  {pages} {165128} (\bibinfo {year} {2012})}\BibitemShut {NoStop}%
\bibitem [{\citenamefont {Dunning}(1971)}]{dunning1971gaussian}%
  \BibitemOpen
  \bibfield  {author} {\bibinfo {author} {\bibfnamefont {T.}~\bibnamefont
  {Dunning}},\ }\href@noop {} {\bibfield  {journal} {\bibinfo  {journal}
  {Journal of Chemical Physics}\ }\textbf {\bibinfo {volume} {55}},\ \bibinfo
  {pages} {716} (\bibinfo {year} {1971})}\BibitemShut {NoStop}%
\bibitem [{\citenamefont {Parker}\ \emph {et~al.}(2013)\citenamefont {Parker},
  \citenamefont {Seideman}, \citenamefont {Ratner},\ and\ \citenamefont
  {Shiozaki}}]{parker2013communication}%
  \BibitemOpen
  \bibfield  {author} {\bibinfo {author} {\bibfnamefont {S.~M.}\ \bibnamefont
  {Parker}}, \bibinfo {author} {\bibfnamefont {T.}~\bibnamefont {Seideman}},
  \bibinfo {author} {\bibfnamefont {M.~A.}\ \bibnamefont {Ratner}}, \ and\
  \bibinfo {author} {\bibfnamefont {T.}~\bibnamefont {Shiozaki}},\ }\href@noop
  {} {\bibfield  {journal} {\bibinfo  {journal} {The Journal of chemical
  physics}\ }\textbf {\bibinfo {volume} {139}},\ \bibinfo {pages} {021108}
  (\bibinfo {year} {2013})}\BibitemShut {NoStop}%
\bibitem [{\citenamefont {Hehre}\ \emph {et~al.}(1972)\citenamefont {Hehre},
  \citenamefont {Ditchfield},\ and\ \citenamefont {Pople}}]{hehre1972self}%
  \BibitemOpen
  \bibfield  {author} {\bibinfo {author} {\bibfnamefont {W.~J.}\ \bibnamefont
  {Hehre}}, \bibinfo {author} {\bibfnamefont {R.}~\bibnamefont {Ditchfield}}, \
  and\ \bibinfo {author} {\bibfnamefont {J.~A.}\ \bibnamefont {Pople}},\
  }\href@noop {} {\bibfield  {journal} {\bibinfo  {journal} {The Journal of
  Chemical Physics}\ }\textbf {\bibinfo {volume} {56}},\ \bibinfo {pages}
  {2257} (\bibinfo {year} {1972})}\BibitemShut {NoStop}%
\end{thebibliography}
\end{document}